\documentclass[journal=jacsat,layout=twocolumn,manuscript=article]{achemso}

\setkeys{acs}{keywords = true}
\SectionNumbersOn
\usepackage[version=3]{mhchem} 
\usepackage{booktabs}%
\usepackage{graphicx}%
\usepackage{multirow}%
\usepackage{chemformula}%
\usepackage{amsmath,amssymb,amsfonts}%
\usepackage{amsthm}%
\usepackage{mathrsfs}%
\usepackage{algorithm}%
\usepackage{algorithmicx}%
\usepackage{algpseudocode}%
\usepackage{listings}%
\usepackage{lmodern}
\usepackage{anyfontsize}
\usepackage[T1]{fontenc}
\usepackage{xcolor}%
\usepackage{textcomp}%
\usepackage{manyfoot}%
\usepackage{bm}
\usepackage{subfiles}

\author{Minseok Moon}

\author{Seungwoo Hwang}

\author{Jaesun Kim}
\author{Yutack Park}
\author{Changho Hong}
\email{mk01071@snu.ac.kr}
\author{Seungwu Han}
\email{hansw@snu.ac.kr}
\phone{+82 (02) 880 1541}

\affiliation[SNU]{%
  Department of Materials Science and Engineering and Research Institute of Advanced Materials,  
  Seoul National University, Seoul 08826, Korea
}

\alsoaffiliation[KIAS]{%
  Korea Institute for Advanced Study, Seoul 02455, Korea
}

\title[]
  {Unveiling defect motifs in amorphous GeSe using machine learning interatomic potentials}

\keywords{ovonic threshold switching, amorphous GeSe, mid-gap defect, machine learning interatomic potential, density functional theory, electronic structure}

\begin{document}






\begin{abstract}

Ovonic threshold switching (OTS) selectors are pivotal in non‑volatile memory devices due to their nonlinear electrical characteristics and polarity-dependent threshold voltages. However, the atomic-scale origins of the defect states responsible for these behaviors remain unclear. In this study, we systematically investigate defects in amorphous GeSe using molecular dynamics simulations accelerated by machine learning interatomic potentials (MLIPs). We first benchmark several MLIP architectures, including descriptor-based potentials and graph neural network (GNN)-based potentials. Our results demonstrate that capturing higher-order interactions, at least four-body correlations, and medium-range structural order is essential for accurately representing amorphous GeSe structures. Our analysis indicates that GNN architectures with multiple interaction layers effectively capture higher-order correlations and medium-range order, thereby preventing spurious defects easily introduced by less descriptive MLIPs. Utilizing the optimized GNN model, we identify two distinct defect motifs across 20 independent 960-atom amorphous GeSe structures: aligned Ge chains associated with defect states near the conduction band, and overcoordinated Ge chains linked to defect states near the valence band. Moreover, we establish correlations between electronic defect levels and specific structural features---namely, the average alignment of bond angles in aligned chains and the degree of local Peierls distortion around overcoordinated Ge atoms. These insights provide a theoretical framework for interpreting experimental observations and deepening the understanding of defect-driven OTS phenomena in amorphous GeSe.

\end{abstract}

\maketitle

\section{INTRODUCTION}\label{sec1}

Non-volatile memory is a critical component in computer storage class systems but faces significant challenges concerning operation speed~\cite{chen2016review}. A notable advancement involves the use of ovonic threshold switching (OTS) materials, characterized by a nonlinear electrical response: the current remains very low until a specific threshold voltage is exceeded, at which point it increases substantially\cite{10.1557/mrs.2019.206}. OTS materials are stacked onto phase-change materials (PCMs) to significantly mitigate sneak currents, thereby improving the reliability of read operations in 3D crossbar arrays used in phase-change memory\cite{raoux2014phase}. Recent studies have indicated that the threshold voltage of amorphous GeSe and SiGeAsSe alloys is polarity-dependent, enabling the development of selector-only memory devices that offer simplified stacking processes and reduced energy consumption by eliminating the need for stacking PCMs\cite{10.1002/pssr.202200417, 10.1109/ted.2023.3252491, clima2020ovonic}. To further enhance device performance, optimizing parameters such as on-state current density, off-state leakage current, threshold voltage, switching speed, and endurance is essential\cite{10.1007/s40820-023-01289-x,10.1088/1361-6641/aa7c25}.


Amorphous GeSe is a prominent material that satisfies critical requirements of high ON current and good thermal stability \cite{10.23919/vlsit.2017.7998207, 10.1109/iedm.2017.8268323}. In amorphous Ge-Se alloys, increasing the Ge content lowers the threshold voltage but compromises thermal stability, exhibiting a performance trade-off \cite{10.1038/s41598-024-57029-7, clima2020ovonic, 10.1063/1.4714705}. To further optimize these alloys, doping strategies—such as incorporating Si or N—have been employed to modulate the threshold voltage and enhance thermal stability \cite{10.1016/j.tsf.2021.138837,clima2020ovonic,10.1063/5.0055861}. Additionally, a composition doped with As and Si is reportedly utilized in the commercialized Intel 3D XPoint\textsuperscript{\textregistered} memory \cite{10.1557/mrs.2019.206}. 

To optimize the electrical performance of OTS, it is essential to first understand the origin of their unique nonlinear response in the amorphous phase. To elucidate the mechanism underlying OTS behavior, numerous theoretical models have been proposed\cite{10.1007/s40820-023-01289-x, 10.1063/1.4738746}. Among these, the Poole-Frenkel model described in Ref.~\citenum{ielmini2007analytical} is noteworthy because it quantitatively captures the full $I$-$V$ characteristics, including subthreshold, switching, and ON-state regimes. This model assumes the presence of localized gap states, attributing conduction in the subthreshold regime to electron transfer via thermal emission between trap states. With increasing applied bias voltage, the potential barrier between defect states decreases approximately in proportion to the product of the electric field and half the intertrap distance, thereby facilitating electron transport. At the threshold voltage, electrons occupy shallow traps or states near the band edges, initiating ballistic tunneling through the conduction band, which leads to the high current characteristic of the ON state. In addition to the Poole-Frenkel model, many other theoretical models similarly invoke localized mid-gap states\cite{10.1007/s40820-023-01289-x, 10.1063/1.4738746,buscemi2023hydrodynamic,emin2006current,degraeve2021modeling,nardone2009unified}. However, the existence and atomic nature of these defects remain controversial due to conflicting experimental evidence. For example, amorphous 
chalcogenides are known for their seemingly contradictory properties, simultaneously exhibiting signatures of both high and low densities of states (DOS) within the band gap \cite{10.1063/1.4738746, 10.1007/978-3-030-69598-9}. Specifically, electron spin resonance (ESR) signals are absent under dark conditions, implying a low DOS of localized states near the Fermi level. However, ESR becomes active upon illumination, indicating a high DOS of gap states that can be optically excited. Optical absorption measurements show that photons with energies near the band gap are absorbed in regions, a sign of a low density of gap states, whereas photoluminescence peaks at approximately half the band gap energy, suggesting a high DOS near the Fermi level. The absence of ESR, the activation of photo-induced ESR, optical absorption, and the photoluminescence behavior have also been reported in Ge$_x$Se$_{1-x}$ alloys \cite{10.1103/physrevb.15.2278, 10.1016/0038-1098(78)90805-0, 10.1080/13642817908246355, 10.1016/0022-3093(80)90321-x, 10.1143/jjap.16.67}.

To address these paradoxes, the valence alternation pair (VAP) model has been proposed \cite{10.1103/physrevlett.37.1504, 10.1063/1.328036}. According to this model, chalcogen atoms typically have two bonds and remain neutral in their ground state. However, the presence of closely spaced defects—one overcoordinated and the other undercoordinated—results in electron redistribution between these defect sites. Consequently, the overcoordinated defect acquires a positive charge, and the undercoordinated defect acquires a negative charge in their respective ground states, rather than remaining neutral. Due to this electron redistribution, the positively charged defect state is unoccupied, while the negatively charged defect state becomes doubly occupied. Upon illumination, these defects revert to neutral configurations characterized by singly occupied states. Although the VAP model remains popular for explaining defect phenomena in amorphous chalcogenides, discrepancies between predicted and experimentally observed defect densities in amorphous Se continue to question the existence and significance of VAPs \cite{10.1007/978-3-030-69598-9}.

Computational studies employing density functional theory (DFT)-based molecular dynamics (MD) simulations offer an alternative viewpoint to the VAP model. One prominent defect structure in amorphous GeSe involves extended Ge chains characterized by homopolar Ge-Ge bonds, typically spanning lengths from 1.5 to 3 nm\cite{xu2022deep, 10.1016/j.mee.2019.110996, 10.1002/elt2.46, 10.1002/aelm.202201224, clima2020ovonic}. This defect state is located near the conduction band and remains unoccupied. It is noteworthy, however, that not all Ge chains necessarily create localized states within the band gap \cite{xu2022deep}. Furthermore, Ref.~\citenum{xu2022deep} posits that a high density of three-fold coordinated Se atoms surrounding Ge chains is essential for defect formation, which contrasts the findings of Ref.~\citenum{10.1002/aelm.202201224} stating that Ge atoms enclosed by three-fold coordinated Se atoms do not produce localized gap states. 
In addition, previous studies have predominantly focused on unoccupied localized defect states near the conduction band. Given the distinctive electronic properties of amorphous chalcogenides, doubly occupied localized defect states near the valence band may also exist, potentially playing a crucial role in their complex electronic behavior. Therefore, extensive configurational sampling through larger-scale simulations is essential to comprehensively capture the diversity of defect motifs and to reliably identify their common structural origins.

Recently, MD simulations accelerated by machine learning interatomic potentials (MLIPs) have demonstrated accuracies comparable to DFT calculations for amorphous PCMs---materials closely related to OTS systems---such as GeTe~\cite{lee2020crystallization, sosso2018understanding, sosso2019harnessing} and Ge–Sb–Te alloys~\cite{mocanu2018modeling, mocanu2020quench, zhou2021structure}. Studies on PCMs have shown that MLIPs can accurately reproduce not only short-range order descriptors, including radial and angular distribution functions, but also medium-range order metrics, particularly ring statistics. However, if an MLIP fails to accurately reproduce DFT results by generating amorphous structures with an excessive abundance of flat, square-shaped, four-membered rings, this discrepancy can significantly accelerate crystallization rates\cite{lee2020crystallization}. Thus, precise characterization of medium-range order—especially the statistics of four-membered rings—is crucial for reliably modeling amorphous chalcogenide systems. Analogous to PCM studies, radial and angular distributions, along with ring statistics, have also been explored in amorphous GeSe$_2$ using an MLIP~\cite{10.1021/acs.jpclett.1c01272}. Nevertheless, investigations of defect states via MLIPs in amorphous Ge–Se alloys remain limited.

In this study, we extensively explore defects in amorphous GeSe through MD simulations accelerated by MLIPs. Initially, we compare results from MD simulations obtained using different MLIP architectures---the Behler-Parinello-type neural network (BPNN)~\cite{behler2007generalized}, the moment tensor potentials (MTP)~\cite{shapeev2016moment}, and SevenNet~\cite{park2024scalable}---to corresponding DFT data. Our comparative analysis indicates that effective MLIPs must incorporate interactions beyond three-body terms and utilize an extended effective cutoff radius for accurate simulations of defects in amorphous GeSe. Using SevenNet, which satisfies these conditions, we performed MD simulations on 20 independent amorphous cells with 960 atoms, identifying two distinct defect types. Specifically, the Ge chains with overcoordinated Ge atom and aligned Ge chains are found to contribute to localized defect states near the valence and conduction band edges, respectively. Furthermore, in overcoordinated Ge chains, the defect energy levels depend on the degree of Peierls distortion at the overcoordinated Ge atoms. In aligned Ge chains, the defect energy levels shift according to the extent of alignment of the Ge–Ge–Ge bond angles. Finally, based on the identified defect motifs, we propose a defect formation mechanism that consistently explains recent experimental observations in amorphous GeSe.

\section{COMPUTATIONAL METHODS}\label{sec2}

\subsection{Construction of training datasets}\label{sec:training_set}

To construct a training dataset suitable for reproducing atomic structures of a-GeSe, we adopt configurations analogous to those used in prior investigations on PCMs\cite{lee2020crystallization, sosso2019harnessing, lee2023generalizable}, which are closely related to OTS materials. The training dataset for the MLIP models comprises crystalline, amorphous, and interface structures, as well as unary structures of Ge and Se, as detailed in Table~\ref{tab:trainset}.

\begin{table*}[htbp]
\caption{Summary of datasets used for training MLIPs}\label{tab:trainset}
\begin{tabular*}{\textwidth}{@{\extracolsep\fill}cccccc}
\toprule
\multirow{2}{*}{Type}
  & Number of   & Temperature      & MD time      & Sampling         & Number of        \\
  & atoms       & (K)              & (ps)         & interval (fs)    & structures       \\
\midrule
Crystal (strained) &   4--8   & —         & —   & —   & 172   \\
Cubic MD           &   216    & 700       & 15  & 120 & 125   \\
Orthorhombic MD    &   72     & 700       & 15  & 120 & 125   \\
Gamma MD           &   64     & 700       & 15  & 120 & 125   \\
Liquid             &   120    & 1050      & 80  & 100 & 800   \\
Quenching          &   120    & 1050–300  & 300 & 300 & 1000  \\
Amorphous          &   120    & 300       & 88  & 100 & 880   \\
Crystal melt       &   64     & 1050      & 40  & 160 & 250   \\
Ge unary MD        &   60     & 1050      & 10  & 80  & 125   \\
Se unary MD        &   60     & 1050      & 10  & 80  & 125   \\
Interface MD       &   120    & 2000      & 12.5& 40  & 312   \\
\bottomrule
\end{tabular*}
\end{table*}

For the crystalline structures, the dataset includes both equilibrium and strained configurations derived from three crystal phases using primitive cells: cubic, orthorhombic, and the recently synthesized gamma phase \cite{lee2021gamma}. Furthermore, to capture local vibrational dynamics, we sampled crystal configurations via NVT MD simulation at 700 K for 15 ps using supercells with the number of atoms specified in Table~\ref{tab:trainset}.

For melt-quench simulations, the initial simulation cell for ab initio molecular dynamics (AIMD) is generated by superheating 120 randomly distributed atoms for 3 ps at 2000 K. The density of the amorphous cells is fixed at 28.9~\AA$^3$/atom, corresponding to an average hydrostatic pressure of approximately 0 kbar during the annealing process. This value closely matches the experimentally reported density of liquid GeSe\cite{ruska1976change}. The superheated structure is held at 1050 K for 20 ps for a liquid state. Subsequently, the liquid GeSe is quenched to 300 K at a rate of $-10$ K/ps and equilibrated at 300 K for an additional 22 ps to yield the amorphous structure. The training set incorporates trajectories starting from the melting phase and excludes those from the superheating phase. Trajectories from four independent simulations are employed to capture a diverse set of amorphous structures.


The melting process of the crystal can provide information on the energy barrier between liquid and crystal, preventing the unrealistically facile crystallization during melt–quench simulations in amorphous chalcogenide systems\cite{lee2020crystallization}.
The simulation is initiated from an FCC structure and performed as a 40 ps MD simulation at 1050 K. 
In addition to suppressing crystallization, we incorporate snapshots from 10 ps MD simulations of liquid unary Ge and Se at 1050 K, as well as from 12.5 ps MD simulations at 2000 K designed to capture intermixing at the interfaces between unary phases, thus inhibiting phase separation into unary phases. The density of both the unary and interface MD simulations is matched to the amorphous density of GeSe. Initial structures for interface MD simulations are generated by joining the final configurations from unary MD runs. Unary cells are constructed with identical lateral ($x$, $y$) lattice parameters, sharing a common in-plane box, which allows us to join them directly along $z$ without further rescaling or rotation. Including these additional configurations helps rectify the ad hoc energy mappings often encountered in multi-component systems and prevents phase separation during liquid-phase simulations~\cite{yoo2019atomic}.


\subsection{Training machine learning interatomic potentials}

In the present work, we employ three distinct MLIP architectures—two descriptor‐type models and one graph‐based model—to assess their capability for reproducing both atomic and electronic structures of amorphous GeSe. Descriptor‐type MLIPs rely on predefined, handcrafted descriptors that encode local geometries within a fixed cutoff, whereas GNN type MLIPs learn representations end‐to‐end by message passing on an atomic graph, enabling flexible modeling of many‐body and medium- to long-range interactions.

The first type of potential is the Behler-Parinello-type neural network (BPNN) \cite{behler2007generalized}, which employs atom-centered symmetry functions to represent the local environment surrounding a center atom and obtains atomic energy through multi-layer perceptron architecture. The second type of potential is the moment tensor potential (MTP) \cite{shapeev2016moment}, which represents atomic contributions in the form of a linear expansion of a set of polynomial basis functions. The basis functions are constructed with all the possible contractions of the moment tensor descriptors. The last type of potential is E(3)-equivariant GNN interatomic potential, which employs E(3)-equivariant graph convolutions for interactions of node features to represent atomic environments\cite{batzner20223, batatia2022mace}. 

We train BPNN potentials using the SIMPLE-NN package \cite{lee2019simple}. The neural network architecture consists of two hidden layers, with the structure of 92-30-30-1 with atom-centered symmetry functions as input features \cite{behler2011atom}. The atom-centered symmetry functions include 20 radial and 72 angular components with a cutoff of 8~\AA{}. Detailed parameters for each symmetry function are explained in Table~S2. We transform symmetry functions by principal component analysis and variances of all components are whitened, which accelerates learning curve convergence. During training, the loss function is defined as the sum of mean squared error (MSE) of energy per atom, atomic forces, and stress tensors, with an L2 regularization value of $10^{-8}$. 
On the other hand, 
MTP models are trained via the MLIP package \cite{shapeev2016moment}. The basis functions consist of moment tensor levels up to 26 and a radial basis size of 8 with a cutoff of 8~\AA{}. During training, the loss function, defined as the sum of MSE of energies, forces, and stresses is optimized.

Finally,
we train E(3)-equivariant GNN interatomic potentials using the SevenNet package \cite{batzner20223}. Node features are updated in interaction layers with the maximum order of spherical harmonics ($l_{\rm max}$) up to 3. The radial function consists of 8 radial Bessel functions with a cutoff of 4~\AA{} passed through a 64-64 fully connected network. The number of channels in the node feature is 32.
E(3)-equivariant GNN potentials incorporating the self-tensor product, characteristic of the MACE-type architecture\cite{batatia2022mace}, are also trained using the SevenNet package. The self-tensor product introduces tensor products of node features at the end of each message-passing layer, enabling the representation of higher-order multi-body interactions without additional message-passing layers. The maximum correlation order, denoted by $\nu$, refers to the number of self-tensor products applied to the node features. We varied the number of convolution layers (up to three) as well as $\nu$ (up to three). In these MACE-type models, the maximum angular momentum channels ($l_{\rm max}$) for node and edge features are set to 2 and 3, respectively.

\subsection{DFT calculations}

All DFT calculations are performed using the Vienna Ab initio Simulation Package (VASP) \cite{kresse1996efficient} with the projector-augmented-wave (PAW) method \cite{blochl1994projector} and the Perdew–Burke–Ernzerhof (PBE) functional \cite{perdew1996generalized}. For AIMD simulations, a plane-wave cutoff energy of 200 eV and the $\Gamma$-point sampling are employed. To enhance the accuracy of the training dataset, total energies, atomic forces, and stress tensors of structures subsampled from trajectories are recalculated using an increased cutoff energy of 300 eV and the Baldereschi point ($\frac{1}{4},\frac{1}{4},\frac{1}{4}$) \cite{baldereschi1973mean}. These computational parameters ensure convergence within 1 meV/atom for total energies, 50 meV/\AA{} for atomic forces, and 1 kbar for pressure.

The DOS of amorphous structures is calculated using geometries optimized by each potential. For these calculations, a cutoff energy of 300 eV is employed, and a $3\times3\times3$ k-point grid is applied to 120-atom cells, whereas the $\Gamma$-point is used for 960-atom cells. Although the PBE functional underestimates the band gap\cite{perdew1983physical}, it reproduces the overall electronic band structure of GeSe\cite{gu2019enhanced}. Therefore, all electronic properties are analyzed using PBE for low computational cost. The inverse participation ratio (IPR) is calculated according to the definition provided in Ref.~\citenum{dong1998atomistic}. The energy scales in the DOS and IPR plots have been aligned by referencing the average electrostatic potential of Se atoms from a single 120-atom DFT amorphous cell\cite{butler2015band, walsh2011multi}. Defect states are defined as localized states within the mobility gap, where the mobility edges $E_{\rm v}$ and $E_{\rm c}$ are determined by fitting the DOS to $(E_{\rm v} - E)^{1/2}$ and $(E - E_{\rm c})^{1/2}$ for the valence and conduction band, respectively\cite{tauc1974optical}.

\section{RESULTS AND DISCUSSION}\label{sec3}

\subsection{Comparative evaluation of MLIP architectures}
To investigate the performance of the MLIP models, we first examine their capability to accurately reproduce amorphous structures, particularly focusing on structural characteristics closely linked to the electronic structure observed in the DOS. As mentioned in the introduction, if an MLIP generates amorphous structures containing an excessive number of flat, square-shaped, four-membered rings, crystallization rates in PCMs become significantly accelerated\cite{lee2020crystallization}. This behavior emphasizes the necessity for MLIPs to accurately represent higher-order correlation beyond three-body terms within their feature vectors. In addition to correctly describing high-order $N$-body interactions, the receptive field of the MLIP must extend beyond the length scale of trap states to faithfully reproduce defect configurations consistently reported in previous literature\cite{xu2022deep, 10.1016/j.mee.2019.110996, 10.1002/elt2.46, 10.1002/aelm.202201224, clima2020ovonic}. To systematically evaluate differences resulting solely from variations in MLIP architecture, identical training datasets are employed across all examined MLIP models.


Before presenting results obtained from different MLIP architectures, we first highlight key distinctions among BPNN, MTP, and SevenNet concerning the construction of $N$-body feature vectors and the receptive field each model can consider. BPNN employs symmetry functions, where $G^2$ represents two-body interactions and $G^4$ encodes three-body interactions within a fixed cutoff radius\cite{behler2011atom}. MTP enriches higher-order features by increasing the rank of moment tensors, explicitly generating higher-order correlations while remaining constrained by a single cutoff radius\cite{shapeev2016moment}. In contrast, SevenNet expands both maximum body order and receptive field simultaneously as the number of interaction layers increases. The MACE-type architecture in SevenNet introduces self-tensor products of node features at the end of each interaction layer, enabling the decoupling of maximum correlation order from receptive field expansion that occurs when additional layers are added~\cite{batatia2022mace}.



\subsubsection{Validation of MLIP models on small-scale amorphous GeSe}\label{sec:validation_small}

Representative hyperparameters for each MLIP are selected to validate how these key distinctions influence the amorphous structure of a-GeSe. The cutoff radius for both BPNN and MTP is fixed at 8~\AA. BPNN captures three-body interactions through the symmetry function $G^4$, whereas MTP extends the correlation order to four-body interactions by employing a moment tensor level up to 26. In contrast, the SevenNet model incorporates four message-passing layers, with a 4~\AA{} cutoff, yielding a receptive field of 16~\AA{} and capturing five-body orders.

The root mean square error (RMSE) values for energy and forces, evaluated on both training and test sets, are summarized in Table~S1. The test set comprises an independent melt-quench AIMD simulation trajectory, distinct from the trajectories employed during training. All models achieve RMSE values on the test set within the ranges of 4.4--9.7 meV/atom for energies and 0.16--0.28 eV/\AA{} for forces, with the SevenNet model exhibiting the lowest errors. Parity plots in Figure~S1 display the correlation between DFT and MLIP-predicted energies and forces for the test set. Despite similarly low overall errors, energies predicted by the BPNN model at the low-energy region, corresponding to amorphous configurations, are systematically shifted upward by approximately 20 meV/atom.

For further validation of atomic and electronic structures, we generate five independent amorphous GeSe structures using each potential by following the melt-quench protocol described in Section~\ref{sec:training_set}. Supplementary Figure~S2 presents radial distribution functions (RDFs) and angular distribution functions (ADFs), averaged over all five trajectories at the annealing temperature. The amorphous structures generated by MTP and SevenNet closely reproduce the main features observed in the RDFs and ADFs calculated by DFT. In contrast, the BPNN model produces a notably sharper peak at 90° in the ADF, as depicted in Supplementary Figure~S2e. This deviation will be analyzed in detail in the subsequent section.

\begin{figure*}[htbp]
  \centering
  \includegraphics[width=170mm]{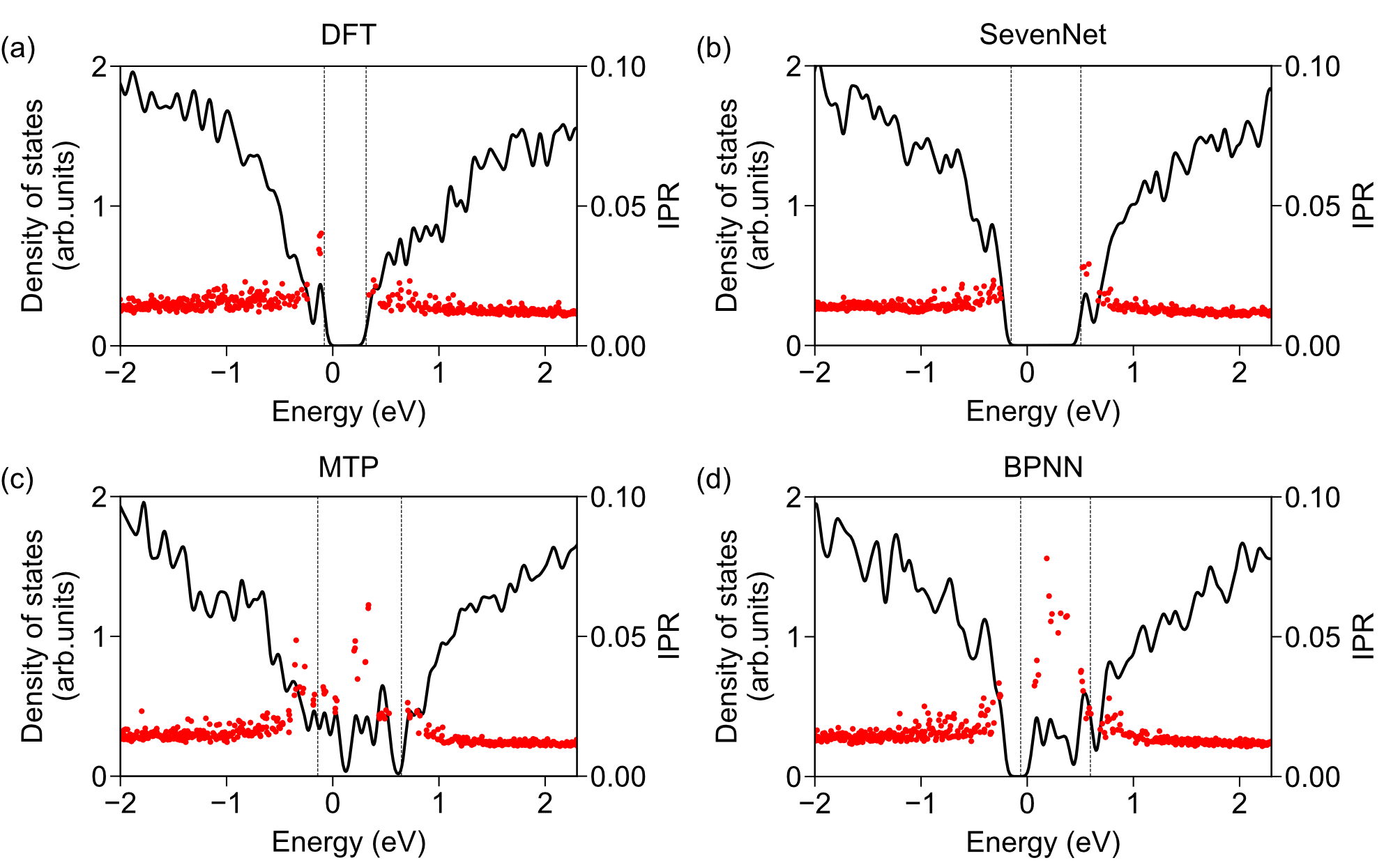}%
  \caption{Electronic DOS (black lines) and IPR (red dots) for amorphous structures generated using (a) DFT, (b) SevenNet, (c) MTP, and (d) BPNN. Dotted lines indicate mobility edges in each DOS plot.}
  \label{fig:DOS}
\end{figure*}

To verify the accurate reproduction of electronic structures, we performed single-point DFT calculations to compute the DOS and IPR for amorphous structures optimized using each potential. Figure~\ref{fig:DOS} compares the electronic DOS and IPR computed from the DFT and each MLIP-derived structure. Structures generated using DFT (Figure~\ref{fig:DOS}a) and SevenNet (Figure~\ref{fig:DOS}b) clearly exhibit distinct band gaps without mid-gap states. In contrast, structures derived from MTP and BPNN (Figure~\ref{fig:DOS}c,d) contain notable mid-gap states, characterized by elevated IPR values indicating increased electronic localization compared to states near the valence and conduction band edges. Additional DOS and IPR data for structures generated using DFT and MLIPs are provided in Figures~S3--S6, consistently supporting these observations.

\begin{figure}[htbp]
  \centering
  \includegraphics[width=84mm]{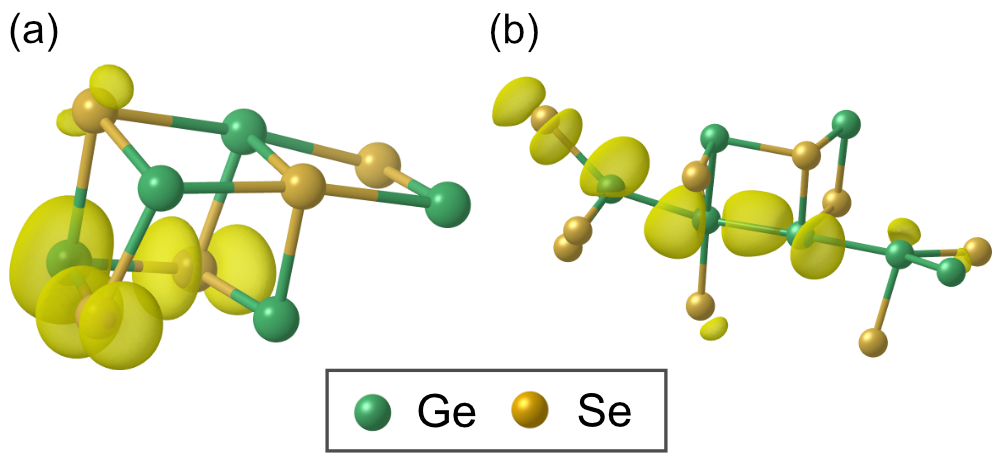}%
  \caption{Partial charge distributions illustrating localized mid-gap states associated with (a) cubic-like motifs from the BPNN model and (b) aligned Ge–Ge chains from the MTP model. Charge density isosurfaces highlight regions of localization. Only atoms contributing to the localized mid-gap charge density are shown.}
  \label{fig:Partial_charge}
\end{figure}

Further analysis of partial charge distributions confirms that the mid-gap states originate from localized defect structures. Figure~\ref{fig:Partial_charge} shows representative atomic configurations associated with mid-gap states, visualized with charge density isosurfaces, from the BPNN and MTP models. In the BPNN-generated structures, mid-gap states localize predominantly around cubic-like structural motifs, whereas in the MTP-generated structures, these states concentrate along aligned Ge chains. The Ge chains in the MTP structures span approximately 8–12~\AA{}, exceeding the cutoff length employed by the MTP. Notably, both defect types persist even after subsequent DFT relaxation of the MLIP-generated structures, highlighting the critical requirement for MLIPs to accurately reproduce atomic structures in order to reliably predict electronic structures. In the following subsections, we explore how the correlation order and effective receptive field of MLIPs influence the formation of spurious defect structures.

\subsubsection{Impact of multi-body correlation order}

\begin{figure}[htbp]
\centering
\includegraphics[width=84mm]{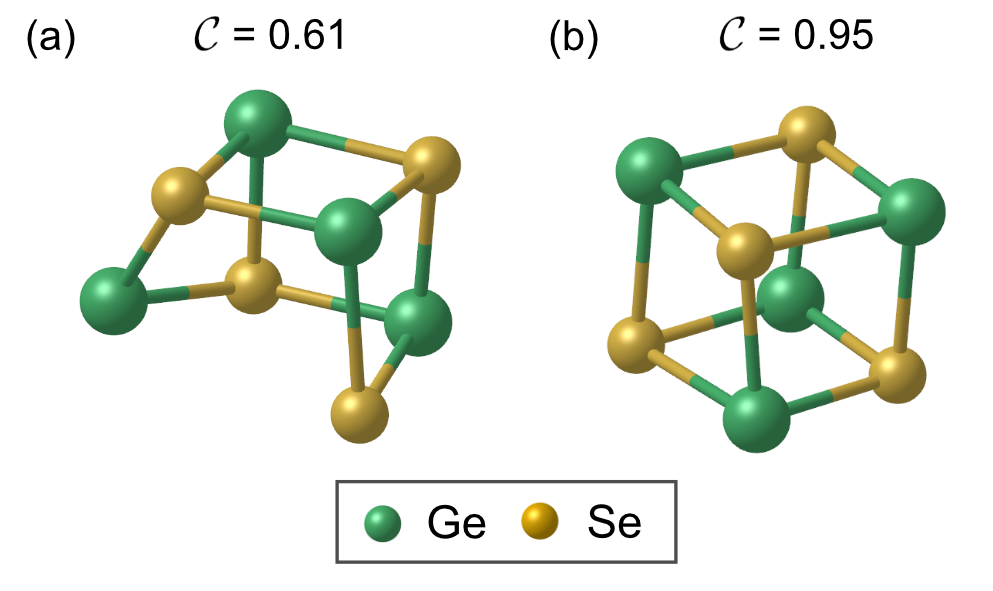}
\caption{Atomic configurations of cubic structures extracted from amorphous GeSe models with (a) low and (b) high cubicity ($\mathcal{C}$).}
\label{fig:Cubic_seed}
\end{figure}

To understand why cubic-like motifs emerge only in lower body-order models, we investigate how the maximum body-order influences the formation of these motifs. Before discussing the details, we define the characteristics of these cubic motifs. These motifs are initially identified by atoms surrounded by three orthogonally arranged square rings. Typical cubic-like motifs are illustrated in Figure~\ref{fig:Cubic_seed}. Some motifs exhibit significant distortion (Figure~\ref{fig:Cubic_seed}a), whereas others exhibit only minor distortion, closely resembling an ideal cubic structure and giving rise to defect states (Figure~\ref{fig:Cubic_seed}b). We note that duplicate structures are removed when counting the number of motifs. To quantify the distortion of these motifs, we define cubicity ($\mathcal{C}$) as:
\begin{equation}
\mathcal{C}
= \frac{1}{1 + \epsilon_\ell^4 + \epsilon_\theta^4}
\end{equation}
where $\epsilon_\ell$ and $\epsilon_\theta$ are the average deviation of bond length and bond angle, respectively. The average deviations are defined as:
\begin{align}
\epsilon_\ell = \frac{1}{12}\frac{\sum_{i=1}^{12} \lvert \ell_i - \bar{\ell}\rvert}{\bar{\ell}},\
\epsilon_\theta =\frac{1}{24}\frac{\sum_{i=1}^{24} \lvert \theta_i - 90\rvert}{90}
\end{align}
where  $\ell_i$ and $\theta_i$ represent each bond length (\AA) and bond angle (degree) within the motif, respectively, and $\bar{\ell}$ denotes the mean bond length of the motif.

\begin{figure}[htbp]
\centering
\includegraphics[width=84mm]{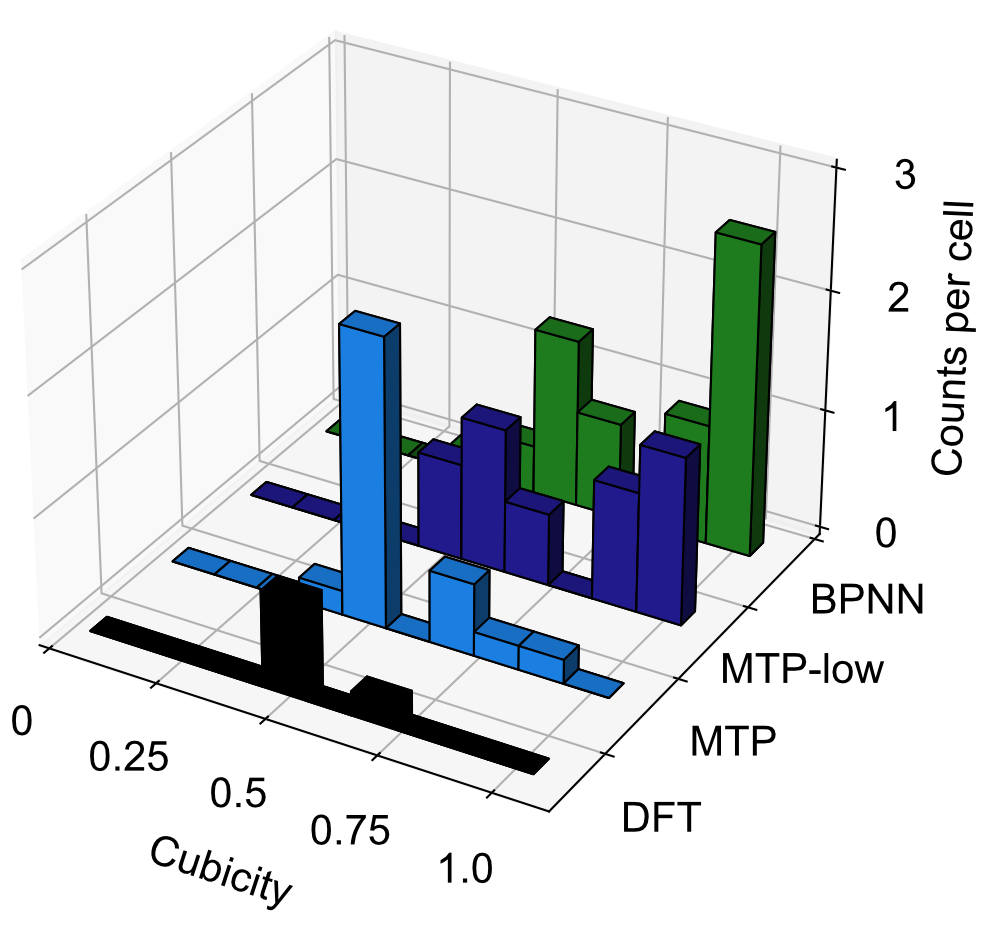}
\caption{Histogram of cubicity distributions for cubic motifs identified in a-GeSe produced by DFT and various descriptor-type MLIPs.}
\label{fig:MTP_lev_cubicity}
\end{figure}

To investigate whether the correlation order of features affects the formation of cubic motifs, we introduce a low-level MTP variant (MTP-low), restricted to three‐body interactions by limiting the maximum moment tensor level to 20. Using the MTP-low model, we generate five independent amorphous cells following the same protocol as employed for the other models (Section~\ref{sec:training_set}). Figure~\ref{fig:MTP_lev_cubicity} compares the cubicity distributions within the amorphous cells generated by different models. High-cubicity motifs are prevalent in the BPNN and MTP-low models but nearly absent in the MTP and DFT structures. Furthermore, the MTP-low model generates localized defects centered around these high-cubicity motifs, closely resembling the behavior observed in the BPNN model. Conversely, the majority of cubic motifs found in the MTP and DFT structures exhibit low cubicity, as illustrated in Figure~\ref{fig:MTP_lev_cubicity}.


Correlation order similarly proves significant in GNN architectures. We demonstrate this by presenting the cubicity distribution obtained from a GNN employing a single interaction layer with self-tensor products, following the MACE-type design\cite{batatia2022mace}. This particular design choice addresses the issue encountered in NequIP-type architectures\cite{batzner20223}, the default model used by SevenNet, wherein increasing the number of interaction layers simultaneously expands both the correlation order and the receptive field. To isolate the effect of correlation order, we restrict our analysis to a single interaction layer with a fixed cutoff distance of 8~\AA{}, and vary the maximum correlation order $\nu$, which determines how many times the self-tensor product is applied, up to $\nu=4$. Thus, the effective body order becomes $(\nu + 1)$ without modifying the receptive field \cite{batatia2025design}. Specifically, we evaluate three configurations: a $\nu=2$ model, capable of capturing up to three-body interactions (similar to BPNN and MTP-low); a $\nu=3$ model, capable of capturing up to four-body interactions (equivalent to MTP); and a $\nu=4$ model, capable of capturing up to five-body interactions.

\begin{figure}
\centering
\includegraphics[width=84mm]{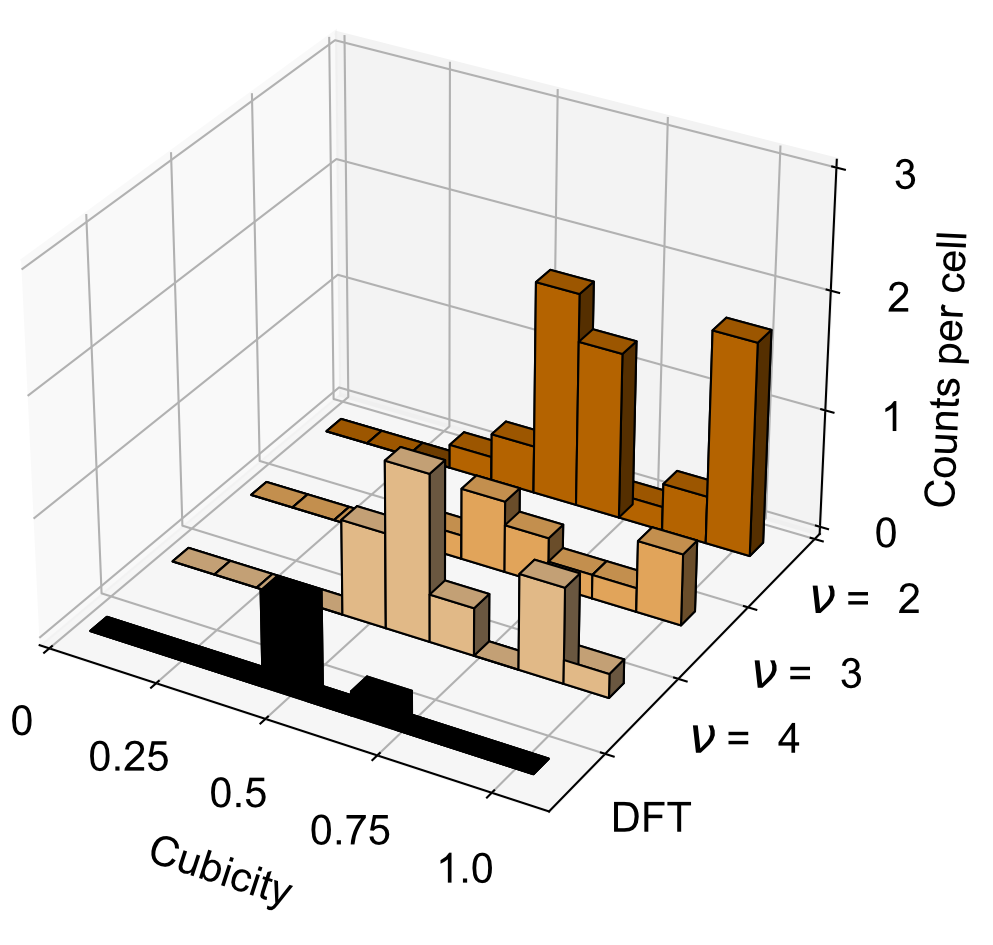}
\caption{Histogram of cubicity distributions for cubic motifs identified in amorphous GeSe structures generated by DFT and single-interaction-layer SevenNet models with varying maximum correlation order ($\nu$).}
\label{fig:STP_cubicity}
\end{figure}

Using the protocol described in Section~\ref{sec:training_set}, five amorphous cells are generated for each SevenNet variant, and the resulting cubicity distributions are presented in Figure~\ref{fig:STP_cubicity}. As shown in Figure~\ref{fig:STP_cubicity}, high-cubicity motifs are prevalent in the \(\nu=2\) model but become significantly suppressed when \(\nu=3\), consistent with the results observed for the MTP and MTP-low models. Further increasing \(\nu\) from 3 to 4 yields negligible additional change, indicating that incorporating four-body interactions alone is sufficient to capture cubic motifs. This observation highlights that models restricted to three-body terms cannot effectively represent the distinctive structural features of cubic-like structures.

\begin{figure*}[htbp]
  \centering
  \includegraphics[width=170mm]{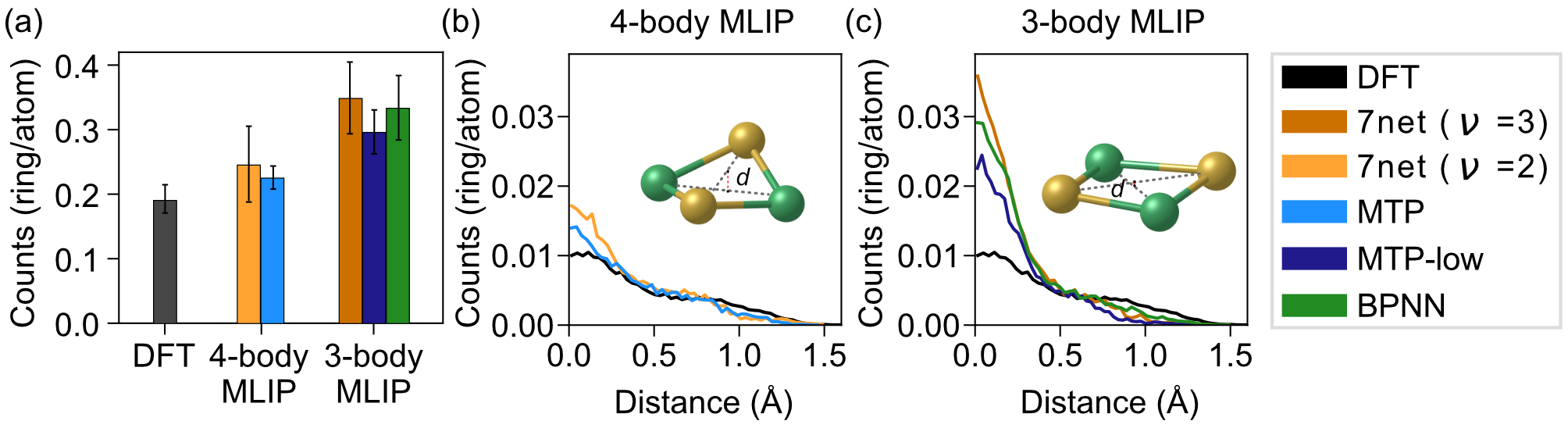}%
  
  \caption{Comparison of square-ring distributions and ring geometry between DFT and MLIP-generated a-GeSe structures. (a) Populations of four-membered rings obtained from DFT, four-body, and three-body order MLIPs. Error bars represent standard deviations from five independent runs. Distributions of inter-diagonal distances within square rings for (b) four-body and (c) three-body MLIPs compared with DFT results.}
  \label{fig:RING}
\end{figure*}

To further explore this behavior, we analyze ring populations using the R.I.N.G.S. code based on King’s criterion\cite{le2010ring}. Figure~\ref{fig:RING}a demonstrates that three-body–limited MLIPs (BPNN, MTP-low, SevenNet ($\nu=2$)) significantly overestimate the population of four-membered rings compared to four-body–inclusive MLIPs (MTP, SevenNet ($\nu=3$)). Analysis of inter-diagonal distances within square rings (Figure~\ref{fig:RING}b,c) confirms that this discrepancy primarily arises from an abundance of flat ring configurations. This excess of flat four-membered rings also manifests as an elevated peak around 90° in the ADF for the BPNN model, as illustrated in Supplementary Figure~S2e. Similar trends are observed for the MTP-low and SevenNet ($\nu=2$) models (results not shown). The excessive occurrence of these flat four-membered rings directly correlates with the formation of artificial cubic motifs and localized defects in a-GeSe structures, which are not observed in DFT-generated cells. In summary, both descriptor-based and GNN-based interatomic potentials confirm that potentials limited to three-body interactions fail to accurately reproduce the structure of a-GeSe, as evidenced by their propensity to form cubic-like defects absent in reference DFT structures.

\subsubsection{Influence of network depth on the effective receptive field}

We demonstrate that MLIPs incorporating correlation orders greater than four-body terms effectively eliminate cubic defects. Nevertheless, these potentials—both descriptor-based and GNN-based—with an 8 \AA{} cutoff radius, still exhibit mid-gap states localized on Ge chain motifs, which are absent in DFT-generated structures. Addressing this limitation necessitates a longer receptive field for the MLIPs. For descriptor-based MLIPs, extending the receptive field by increasing the cutoff radius results in computational complexity that scales approximately as $O(r^3)$, imposing impractical computational costs. For instance, the MTP model training with a 16 \AA{} cutoff radius is estimated to require approximately 40 days on a single Intel Xeon\textsuperscript{\textregistered} 6226R CPU. In contrast, the receptive field in GNN-based MLIPs can be expanded by adding interaction layers, resulting in approximately linear scaling with the number of layers. This favorable scalability allows efficient modeling of defects with extended-range correlations.

Apart from scalability, we emphasize in this section that the effective receptive field of GNNs does not simply scale with the nominal receptive field, defined as the product of the cutoff length and the number of interaction layers. Instead, we observe that it is also strongly influenced by network depth. To demonstrate this, we trained three SevenNet models, each with an identical total receptive field of 16 \AA{}: a four-layer model with a 4 \AA{} per-layer cutoff (c4i4), a two-layer model with an 8 \AA{} per-layer cutoff (c8i2), and a single-layer model with a 16 \AA{} cutoff (c16i1). To ensure consistent coverage of at least four-body interactions across all models, a self-tensor product step is incorporated into the c8i2 and c16i1 models with correlation orders of $\nu=2$ and $\nu=3$, respectively.

\begin{figure*}[htbp]
  \centering
  \includegraphics[width=170mm]{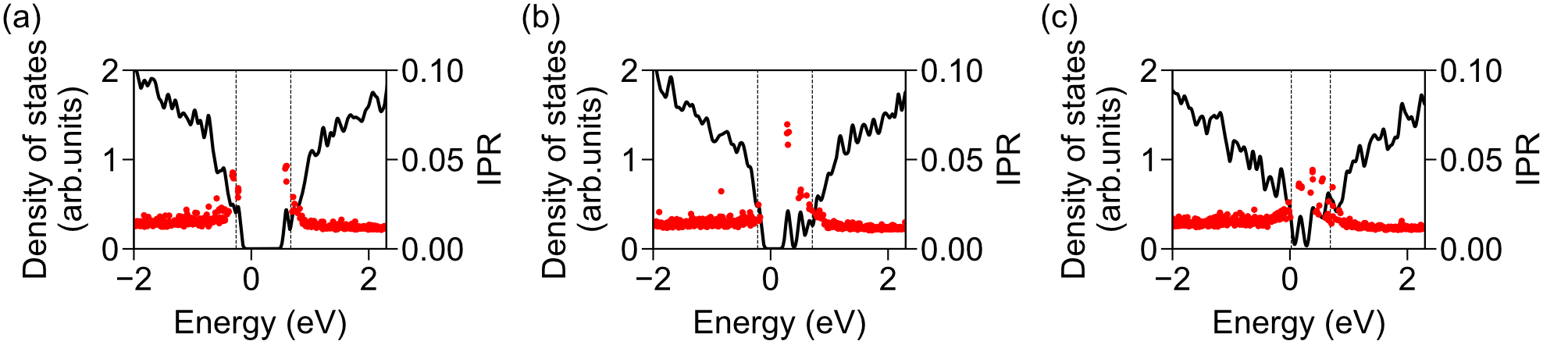}%
  
  \caption{Electronic DOS (black lines) and IPR (red dots) of amorphous structures generated using different numbers of interaction layers: (a) a four-layer model with a 4~\AA{} per-layer cutoff (c4i4), (b) a two-layer model with an 8~\AA{} per-layer cutoff (c8i2), and (c) a single-layer model with a 16~\AA{} cutoff (c16i1). Dotted lines indicate the mobility gap in each DOS plot.}
  \label{fig:DOS_STP}
\end{figure*}

Following the protocol outlined in Section~\ref{sec:training_set}, five amorphous cells are generated for each SevenNet variant. Representative DOS and IPR results for each variant are compared in Figure~\ref{fig:STP_cubicity}. The c4i4 model accurately reproduces the expected band gap (Figure~\ref{fig:STP_cubicity}a), whereas the c8i2 and c16i1 models exhibit pronounced mid-gap states (Figure~\ref{fig:STP_cubicity}b,c). This demonstrates that merely enlarging the cutoff radius in shallow GNN architectures fails to suppress Ge chain defects.

This indicates that the node feature of an atom fails to capture differences arising from distant atomic displacements. To quantify this effect, we adopt the sensitivity metric between nodes separated by multiple hops in GNNs, introduced by Ref.~\citenum{topping2021understanding}:
\begin{equation}
  S_{(u,v)}
  = \max_{i}(
    \frac{\bigl|\Delta \textbf{f}_v\bigr|}
         {\bigl| \Delta \textbf{x}_u^{i}\bigr|}\,)
\end{equation}
Here, $\Delta\textbf{f}_v$ represents the difference vector of the atomic feature vector of atom $v$, computed at the output of the final convolution layer, resulting from a displacement applied to atom $u$. $\Delta \textbf{x}_u^{i}$ denotes the displacement vector applied to atom $u$, and $i$ indexes the displacement directions. We calculate $S{(u,v)}$ by displacing each atom by 0.01~\AA{} in ten random directions, uniformly sampled over the unit sphere, and measuring the resulting changes in atomic features. This metric quantifies how strongly the displacement of atom $u$ influences the feature vector of atom $v$. A low $S_{(u,v)}$ value means that the change in node feature $\mathbf{f}_v$ induced by $\Delta \mathbf{x}_u$ is small, indicating a loss of information\cite{topping2021understanding}.


\begin{figure}[htbp]
  \centering
  \includegraphics[width=84mm]{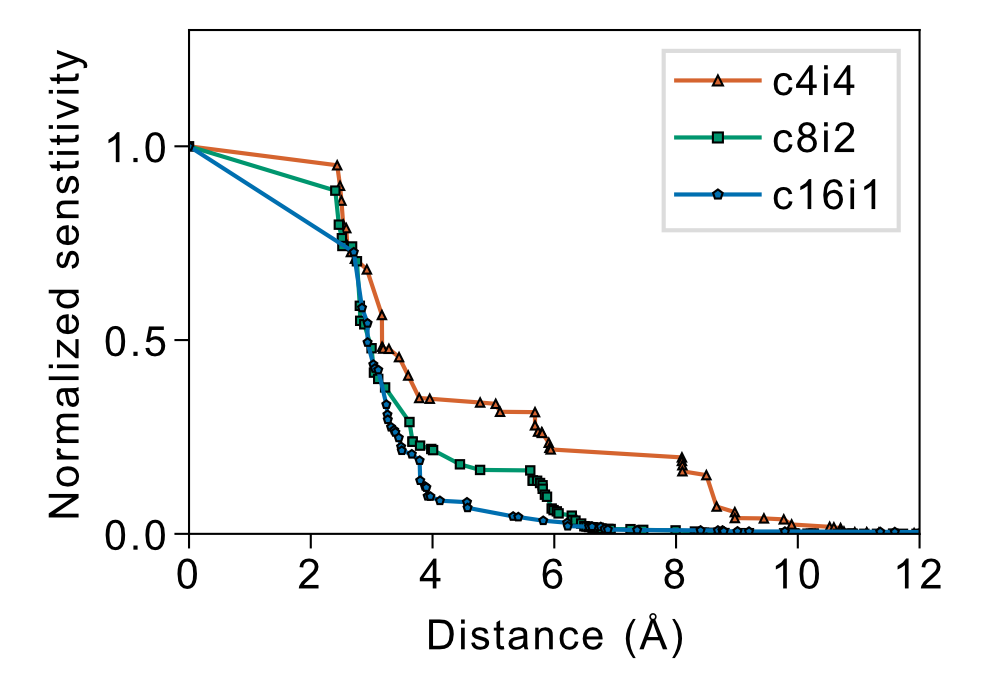}%
  
  \caption{Distance dependence of the normalized sensitivity metric $S_{(u,v)}$. Curves labeled c\textit{X}i\textit{Y} denote models with an \textit{X}~\AA{} cutoff and \textit{Y} interaction layers.}
  \label{fig:Main_sensitivity}
\end{figure}

Sensitivity can vary between atom pairs even at identical interatomic distances. Here, we focus on how sensitivity decays with increasing atomic separation. Therefore, in Figure~\ref{fig:Main_sensitivity}, we plot the maximum normalized sensitivity at each interatomic distance. Sensitivity values are computed for every atom pair, using the minimum image convention, within the amorphous structures and normalized by the sensitivity of an atom pair where the displaced atom and the measured atom are identical. Figure~\ref{fig:Main_sensitivity} clearly illustrates that sensitivity decays before reaching the theoretical receptive field of 16 \AA. Nevertheless, the four-layer model sustain higher sensitivity into medium-range distances (8--10~\AA), whereas sensitivity in the shallower-layer models sharply diminishes around 4--6 \AA{}. This behavior aligns with the presence of defects observed exclusively in the two-layer and single-layer models (Figure~\ref{fig:DOS_STP}b,c).

The rapid sensitivity decay observed for the single-layer model suggests that merely increasing the cutoff radius in descriptor-based potentials is also insufficient for meaningfully extending their effective interaction range. Instead, deeper GNN architectures more effectively capture medium-range atomic correlations. However, further increasing the network depth beyond four layers (up to six layers) does not yield additional sensitivity improvements, as illustrated in Supplementary Figure~S7.

\subsection{Defect structures in amorphous GeSe}

In the preceding section, we demonstrate that accurately modeling defect structures necessitates incorporating at least four-body interaction terms along with a sufficiently long effective receptive field. Consequently, we employ the previously validated SevenNet model described in Section~\ref{sec:validation_small} to generate large-scale amorphous structures and systematically analyze their defect characteristics. Using the same melt-quench method described earlier, but with an increased cell size, we produced 20 independent amorphous structures. Each structure contains 960 atoms within dimensions of approximately $3 \times 3 \times 3$ nm$^3$, sufficiently large to encompass the Ge chain defects reported in Ref.~\citenum{clima2020ovonic}.

Prior to detailed defect characterization, we first verify whether the trained model maintains similar accuracy despite the increased system size. Snapshots extracted from the MLIP melt-quench trajectories are evaluated for their energy and force prediction errors. The RMSE values for energy and force are found to be 5.8 meV/atom and 0.1 eV/\AA, respectively, comparable to errors obtained from the test set of smaller cells. This consistency confirms the reliability of the model for larger systems. Subsequently, electronic structures are computed based on configurations optimized with the PBE functional to mitigate potential artifacts arising from unrelaxed MLIP-generated amorphous structures. Resulting electronic properties, including DOS and IPR values, are presented in Supplementary Figures~S8--S10. The calculated average mobility gap of a-GeSe is 0.7 eV, which is lower than the experimentally determined optical band gap of 1.1 eV\cite{kim2018optical,vaughn2012colloidal}. This underestimation is a known limitation inherent in standard DFT functionals\cite{perdew1983physical}.

\begin{figure*}[htbp]
  \centering
  \includegraphics[width=170mm]{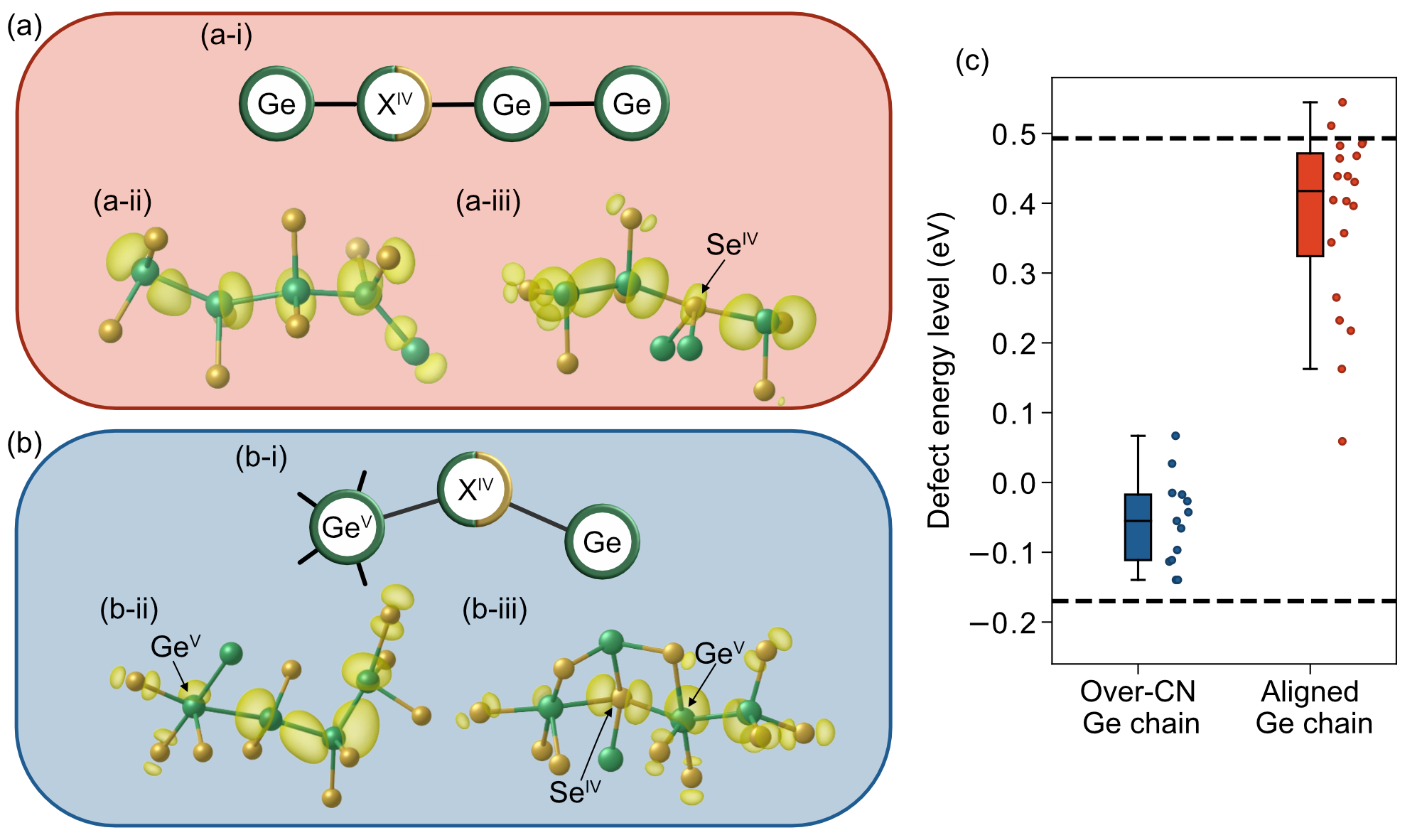}%
  
\caption{Schematics, partial charge distributions, and defect energy levels illustrating two distinct defect motifs identified in a-GeSe: (a) aligned Ge chains associated with defect states near the conduction band, and (b) overcoordinated Ge chains associated with defect states near the valence band. (a-i, b-i) Schematic representations of each motif. (a-ii, b-ii) Partial charge densities corresponding to the aligned and overcoordinated Ge chain defect states, respectively. (a-iii, b-iii) Examples showing the presence of four-coordinated Se atoms (labeled as Se$^\mathrm{IV}$) within both motifs. For partial charge distributions only atoms relevant to the localized charge density are presented. (c) Box plot comparing defect energy levels of the overcoordinated (Over-CN) and aligned Ge chain motifs found in 20 independent 960-atom amorphous cells; dashed horizontal lines represent the mean mobility gaps.}
  \label{fig:Large_defect}
\end{figure*}

By analyzing partial charges and identifying common local atomic environments associated with defects, we categorize the localized defect states into two distinct types: aligned Ge chains near the conduction band and overcoordinated Ge chains near the valence band, as illustrated in Figure~\ref{fig:Large_defect}a,b. Aligned Ge chains are defined as chains comprising at least four Ge atoms, with bond angles between adjacent Ge atoms exceeding 120 degrees. An example of an aligned Ge chain and the partial charge density associated with the defect state is shown in Figure~\ref{fig:Large_defect}(a-ii). Overcoordinated Ge chains consist of at least three Ge atoms, with at least one atom exhibiting five-fold coordination, forming square pyramidal motifs. An example of such an overcoordinated Ge chain, along with the partial charge density associated with the localized state, is depicted in Figure~\ref{fig:Large_defect}(b-ii). For both types of defects, partial charge density reveals strong localization along the Ge chains.

In both defect types, we observe that a four-coordinated Se atom can substitute for a Ge atom to form a defect state. In the schematic depictions provided in Figure~\ref{fig:Large_defect}(a-i) and (b-i), X$^{\rm IV}$ denotes a four-coordinated atom, either Ge or Se. We consistently find that four-coordinated Se atoms constitute defect centers when their Ge–Se–Ge bond angles exceed 150°. Examples of Se incorporation and corresponding partial charge densities are illustrated in Figure~\ref{fig:Large_defect}(a-iii) and (b-iii).

Figure~\ref{fig:Large_defect}c illustrates that defect energy levels of aligned Ge chains and overcoordinated Ge chains are distributed near the conduction and valence bands, respectively. Notably, each defect type exhibits energy level distributions extending approximately 0.3 eV into the band gap. Ge chains lacking alignment or overcoordinated Ge chains exhibiting significant distortion at the overcoordinated Ge atom do not form discrete defect states. Instead, they contribute to tail states in the conduction and valence bands, respectively, as illustrated in Supplementary Figure~S11. This observation provides an explanation for the previously reported finding that not all Ge chains generate defect states \cite{xu2022deep}.

According to the VAP theory described in Ref.~\citenum{clima2020ovonic}, undercoordinated Ge atoms produce defect states near the valence band, whereas overcoordinated Ge atoms introduce states near the conduction band. However, our results for Ge chain motifs contradict this assertion. Specifically, we observe that overcoordinated Ge chains predominantly give rise to defect levels at the valence-band edge, whereas undercoordinated Ge atoms itself does not contribute to trap states. In addition to the local VAP theory focusing on individual atoms, Ref.~\citenum{xu2022deep} proposed the global 8$-$$N$ rule for Ge chains. According to this rule, four-coordinated Ge atoms terminated by two-coordinated Se atoms do not induce defect states; instead, overcoordinated Se atoms (three- or four-fold) introduce excess electrons, creating localized defect states. Nevertheless, our data on Ge chain motifs deviate from this prediction. We find mid-gap defect states in chains terminated by two-coordinated Se atoms (Supplementary Figure~S12a), while chains exclusively capped by three-coordinated Se atoms exhibit no mid-gap states (Supplementary Figure~S12b).

\begin{figure}[!hbtp]
\centering
\includegraphics[width=84mm]{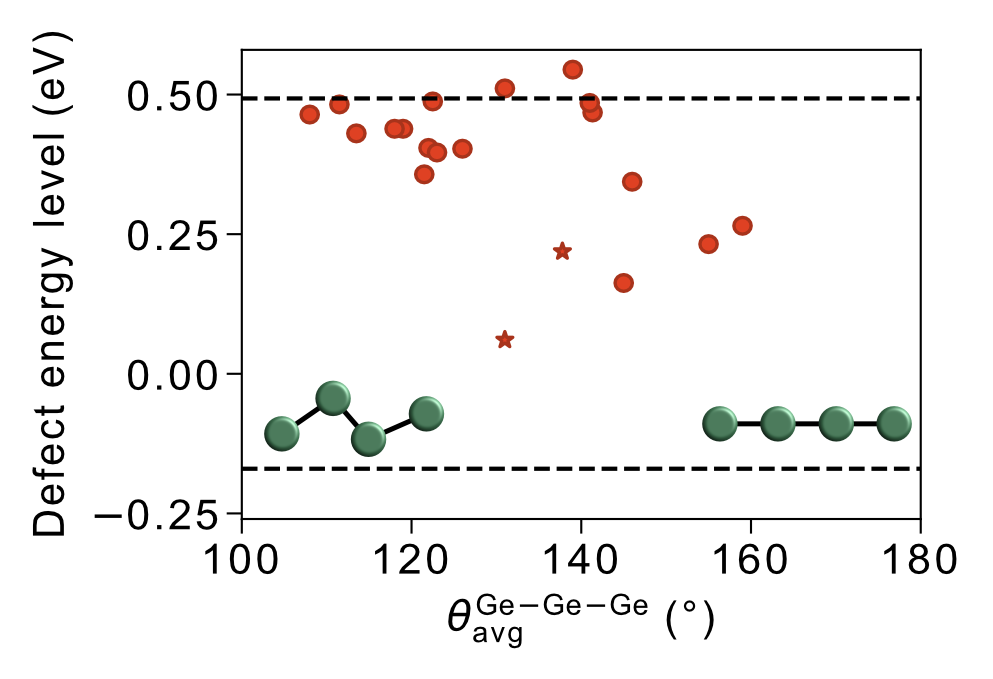}
\caption{Scatter plot illustrating defect energy levels of aligned Ge chains as a function of their average Ge–Ge–Ge bond angles ($\theta_{\mathrm{avg}}^{\mathrm{Ge-Ge-Ge}}$). For motifs containing four-coordinated Se atoms, Ge–Se–Ge angles are included. Stars denote seven-atom chains. Dashed horizontal lines indicate the mean mobility edges. Schematics depict representative structural motifs associated with distinct bond-angle regimes.}
\label{fig:angle_defect}
\end{figure}

To further elucidate the factors influencing each type of defect level found in this study, we examine geometric features associated with the defect levels. First, Figure~\ref{fig:angle_defect} illustrates the correlation between the defect levels of aligned Ge chains and the average Ge–Ge–Ge bond angle along the chain members ($\theta_{\rm avg}^{\text{Ge-Ge-Ge}}$). The defect level deepens with increasing $\theta_{\rm avg}^{\text{Ge-Ge-Ge}}$. Additionally, we observe that Ge chains with relatively small $\theta_{\rm avg}^{\text{Ge-Ge-Ge}}$ occupy tail states near the conduction band (See Supplementary Figure~S11a,b). Beyond bond angles, we find that the number of atoms forming the chain also significantly influences the defect levels. Specifically, points marked with stars in Figure~\ref{fig:angle_defect} correspond to seven-atom chains; these longer chains notably exhibit deeper defect levels compared to shorter chains, even when their average bond angles are similar. Excluding these two cases, the typical chain length observed is four or five atoms.

\begin{figure}[!hbtp]
\centering
\includegraphics[width=84mm]{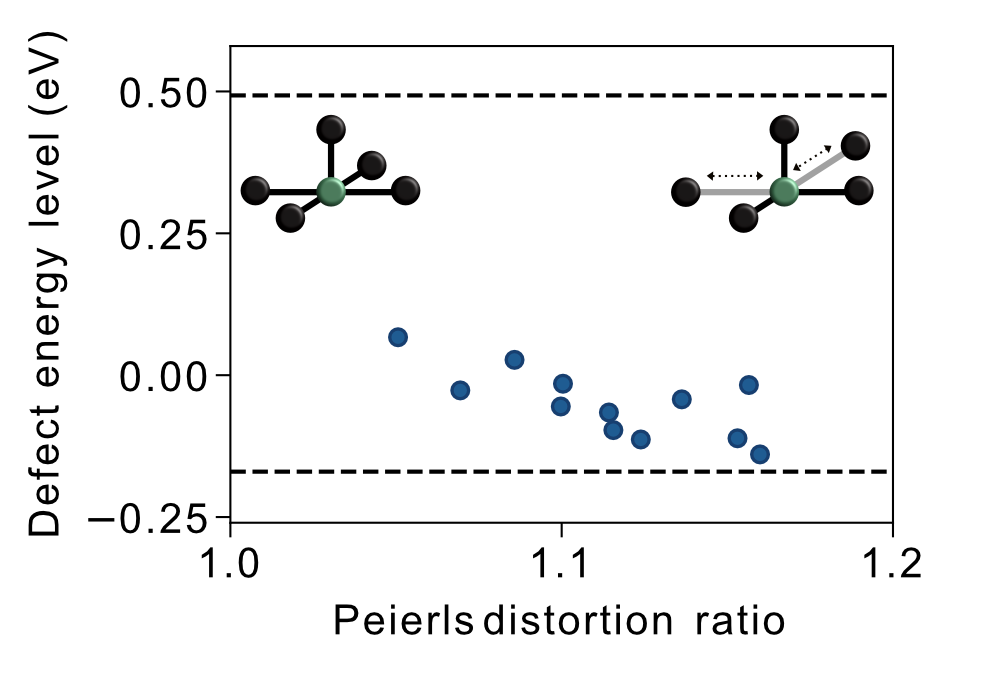}
\caption{Scatter plot illustrating defect energy levels of overcoordinated Ge chains as a function of the Peierls distortion ratio calculated around the overcoordinated Ge atom. Dashed horizontal lines indicate the mean mobility edges. Schematics depict representative local structural motifs corresponding to low (square pyramidal) and high (distorted pyramidal) Peierls distortion ratios.}
\label{fig:length_defect}
\end{figure}

While the average bond angle influences the defect level of aligned Ge chains near the conduction band, we observe that the Peierls distortion ratio around overcoordinated Ge atoms significantly affects the defect levels of overcoordinated Ge chains. The local configurations of overcoordinated Ge atoms that form gap states predominantly adopt a square pyramidal motif, with only one case exhibiting an octahedral motif. We define the Peierls distortion ratio as the ratio of bond lengths between aligned bond pairs; hence, the ratio for the square pyramidal configuration is the average of two distinct aligned bond pairs, whereas it is the average of three aligned bond pairs in the octahedral case. A higher distortion ratio indicates greater distances between some neighboring atoms and the central Ge atom, corresponding to reduced overcoordination. Figure~\ref{fig:length_defect} illustrates the correlation between the distortion ratio and the defect energy level within the mobility gap. As the distortion ratio increases, the defect levels progressively shift toward the band-tail region. Notably, overcoordinated Ge chains with distortion ratios exceeding approximately 1.15 do not contribute to gap states but instead contribute to tail states in the valence band ({Supplementary Figure~S11c,d). Such elevated distortion ratios closely resemble local motifs found in orthorhombic crystalline structures, which exhibit distortion ratios of around 1.3 and are typically regarded as three-fold coordinated at this distortion level.

The geometric characteristics of both Ge chain defects make them strong candidates for elucidating the microscopic origins of the OTS phenomenon. Deep-level transient spectroscopy (DLTS) studies indicate that trap characteristics in a-GeSe deviate significantly from those typical of point defects, exhibiting slow logarithmic trap filling and broad DLTS peaks\cite{hsu2020deep}. Specifically, Ref.~\citenum{hsu2020deep} emphasizes that a broad DLTS peak implies a wide distribution of defect energy levels, consistent with our proposed defect model, in which variations in Ge–Ge–Ge bond angles and Peierls distortion ratios yield diverse defect energy distributions. Furthermore, Ref.~\citenum{10.1109/ted.2022.3169118} reports a significant reduction in threshold voltage during pulses at millisecond scales or shorter, indicating rapid atomic-scale rearrangements as integral to the OTS mechanism. We attribute this threshold voltage reduction to subtle geometric modifications of bond angles and Peierls distortion ratios occurring during ON states and their subsequent recovery.

In addition to Ge chain defects, we propose that initially non-defective Ge chains, characterized by low alignment or high distortion ratios and thus lacking defect states, play crucial roles in our defect model.
When these chains undergo alignment or become less distorted, they account for previously observed experimental phenomena related to defect-state evolution during and after threshold switching. Specifically, inverse photoemission spectroscopy (IPES) measurements demonstrate evolving slopes at the onset of internal photoemission spectra with increasing bias, indicating a gradual shift of band-tail states into the mobility gap \cite{10.1002/pssr.201900672}. Moreover, DLTS measurements conducted following forward bias (SET states) and reverse bias (RESET states) reveal distinct distinct distributions of defect levels for each case \cite{sung2024microscopic}. 
Combined findings from IPES and DLTS experiments suggest that persistent defect states originating from band tails remain stable after bias removal but revert to tail states when reverse bias is applied. According to our model, initially non-defective Ge chains reversibly transform into defect states via subtle adjustments in Ge–Ge–Ge bond angles or distortion ratios, highlighting the sensitivity of electronic states in a-GeSe to minor structural rearrangements. 


We have demonstrated that aligned and overcoordinated Ge chain motifs constitute the primary mid-gap defect centers in amorphous GeSe. In particular, correlations have been established between their geometric characteristics, specifically bond-angle alignment and Peierls distortion ratios, and the resulting defect energy levels. This computational approach leveraging the MLIP can be systematically extended to investigate how variations in Ge:Se stoichiometry or the introduction of specific dopants influence defect structures and their associated energy distributions. Beyond computational predictions, experimental validation of these proposed defect structures remains essential. Notably, the recent observation of OTS induced by bond alignment in GeTe$_x$ via ultrafast THz spectroscopy\cite{seong2025transient} provides a promising method to confirm the proposed origins of defects in amorphous GeSe. An integrated computational and experimental strategy will thus facilitate the materials engineering of OTS materials critical for the development of next-generation nonvolatile memory technologies.


\section{Conclusion}

In this study, we systematically investigated the necessary characteristics of MLIP architectures required for accurately reproducing the atomic structures of amorphous GeSe and identified defect motifs on a large scale. By benchmarking descriptor-based (BPNN, MTP) and GNN-based (SevenNet) potentials against DFT-generated reference data, we demonstrated that capturing higher-order interactions beyond three-body correlations and employing an extended effective receptive field are crucial to avoid artificial defects. Our analysis revealed that lower-order potentials, restricted to three-body terms, frequently produce artificial cubic motifs characterized by an excessive presence of flat, square-shaped, four-membered rings, consequently leading to erroneous mid-gap states. In contrast, potentials capable of describing interactions of at least four-body terms substantially mitigate these artifacts. Additionally, we showed that in GNN-based MLIPs, the effective receptive field critically depends on network depth rather than exclusively on the nominal receptive field size; deeper networks more effectively capture long-range interactions, thus suppressing defects resulting from inadequate medium-range correlations.

Using an optimized SevenNet model incorporating these critical features, we identified two distinct defect motifs in large-scale MD simulations of amorphous GeSe structures: aligned Ge chains associated with conduction-band defect states, and overcoordinated Ge chains linked to valence-band defect states. Furthermore, we correlated specific geometric characteristics---bond-angle alignment and local Peierls distortion---with their corresponding electronic defect levels. This correlation provides deeper insights into the atomic origins of defect states and can facilitate a better understanding and interpretation of experimental observations of amorphous chalcogenide-based materials.


\section*{Supporting Information Available}
The following files are available free of charge.

Supporting Information: This file includes RMSE errors of energy, force, and stress for the training and test sets; parity plots comparing DFT reference values against MLIP models; partial RDFs and ADFs comparing DFT and MLIP models; DOS and IPR for amorphous configurations generated directly by DFT MLIPs; the full set of parameters used in the atom-centered symmetry functions; DOS and IPR results for large-scale amorphous structures generated by SevenNet; Partial charge distribution of valence-band tail and conduction-band tail states.

\section*{Acknowledgement}
This work was supported by Samsung Electronics Co., Ltd(IO201214-08143-01) and the Virtual Engineering Platform Project of the Ministry of Trade, Industry and Energy (MOTIE) of Korea (grant number: P0022336-G02F09901883413-10054408). The computations were carried out at the Korea Institute of Science and Technology Information (KISTI) National Supercomputing Center (KSC-2025-CRE-0110) and the Center for Advanced Computations (CAC) at Korea Institute for Advanced Study (KIAS).

\bibliography{references}


\end{document}